# Self-Organized Periodic Lattices of Chaotic Defects

Frederick H. Willeboordse* and Kunihiko Kaneko†
*Department of Pure and Applied Sciences, University of Tokyo, Komaba, Tokyo 153, Japan*

A novel type of self-organized lattice in which chaotic defects are arranged periodically is reported for a coupled map model of open flow. We find that temporally chaotic defects are followed by spatial relaxation to an almost periodic state when suddenly a next defect appears. The distance between successive defects is found to be generally predetermined and diverging logarithmically when approaching a certain critical point. The phenomena are analyzed and shown to be explicable as the results of a boundary crisis for the spatially extended system.

For their extremely rich phenomenology and their far reaching significance in the study of universal properties in complex systems, coupled map lattices (CML) have drawn much interest in recent years. Among the most widely studied CMLs are lattices of coupled logistic maps since these are of fundamental importance for gaining and deepening the understanding of spatially extended dynamical systems [1–8].

In this letter, we investigate the dynamics of a one-way coupled logistic lattice (OCLL), which is conceptually closely related to open flow systems and therefore relevant for the general study of turbulence (e.g. in technological applications like jet engines), pipe flow and data traffic. In such systems, performance may critically depend on non-linear effects and consequently, it is of great importance to obtain insights into the underlying universal mechanisms. Indeed, we believe that some of our findings may experimentally be verified, like possibly in Ohtsuka and Ikeda's optical system with distributed non-linear elements [9] or pipe flow [10].

In a previous paper [11], we reported the discovery of spatial chaos with temporal periodicity and analyzed the stability of the spatial patterns with a (newly introduced) spatial map which enabled us to predict certain types of non-trivial down-flow behavior. In principle, our analysis was restricted to large coupling strengths though. In the present letter, we will investigate small coupling strengths which contrary to larger ones yield extremely rich temporal dynamics. We report the discovery of a fascinating new type of self-organization in which a lattice is formed by evenly spaced chaotic defects. It is shown that this phenomenon can completely be explained by considering a new low-dimensional map employed here for the first time, and that it is related to a boundary crisis.

The model under investigation can be expressed as,

$$x^i_{n+1} = (1-\epsilon)f(x^i_n) + \epsilon f(x^{i-1}_n), \tag{1}$$

where $n$ is the discrete time, $i$ the discrete space, $\epsilon$ the coupling constant, and $f$ the logistic map which has the nonlinearity $\alpha$ as its parameter [4]. We choose a fixed boundary condition set to 1.0 throughout, but the exact value does not influence the results in a qualitative way.

A rough phase diagram indicating the main patterns relevant to this letter is given in Fig. 1.

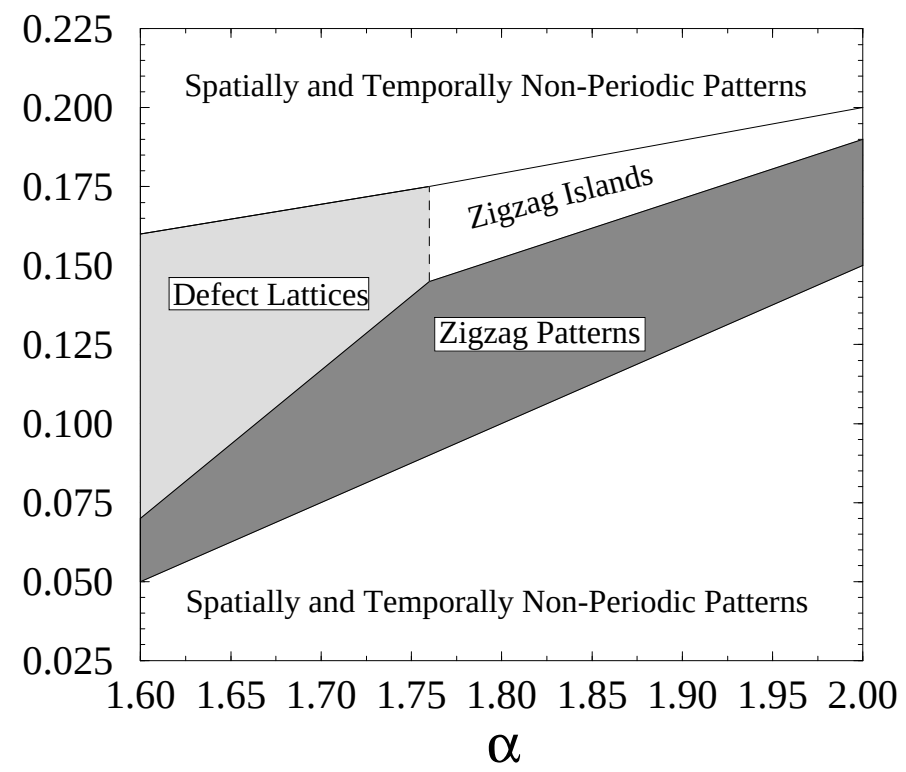

FIG. 1. Rough phase diagram for Eq. 1. The lines are only intended for indicating the approximate region in which the patterns occur. The exact boundaries may depend on both the boundary and initial conditions.

The lower grey area stretching from $\alpha = 1.6$ to $\alpha = 2.0$ is the so-called zigzag region which, mostly except for the lower boundary, has a spatial and temporal periodicity of two. The two temporal phases are furthermore spatially exactly out of phase and consequently there are only two fixed points $x^*_1$ and $x^*_2$ that can easily be found as

$$x^*_{1,2} = \frac{1 \pm \sqrt{4(1-2\epsilon)^2\alpha + 4\epsilon - 3}}{2(1-2\epsilon)\alpha}. \tag{2}$$

For $\alpha \gtrsim 1.76$ we observed the occurrence of coherent zigzag structures within the chaotic sea which are like islands that appear, disappear, grow and shrink. Hence the corresponding area is marked as 'Zigzag Islands'.

Our main discovery here is that for $\alpha \lesssim 1.76$, above the zigzag regime, lattices emerge with evenly spaced defects whose internal dynamics is chaotic when starting from random initial conditions. An example is shown in Fig. 2, where the distance between successive defects is 20 lattice sites.

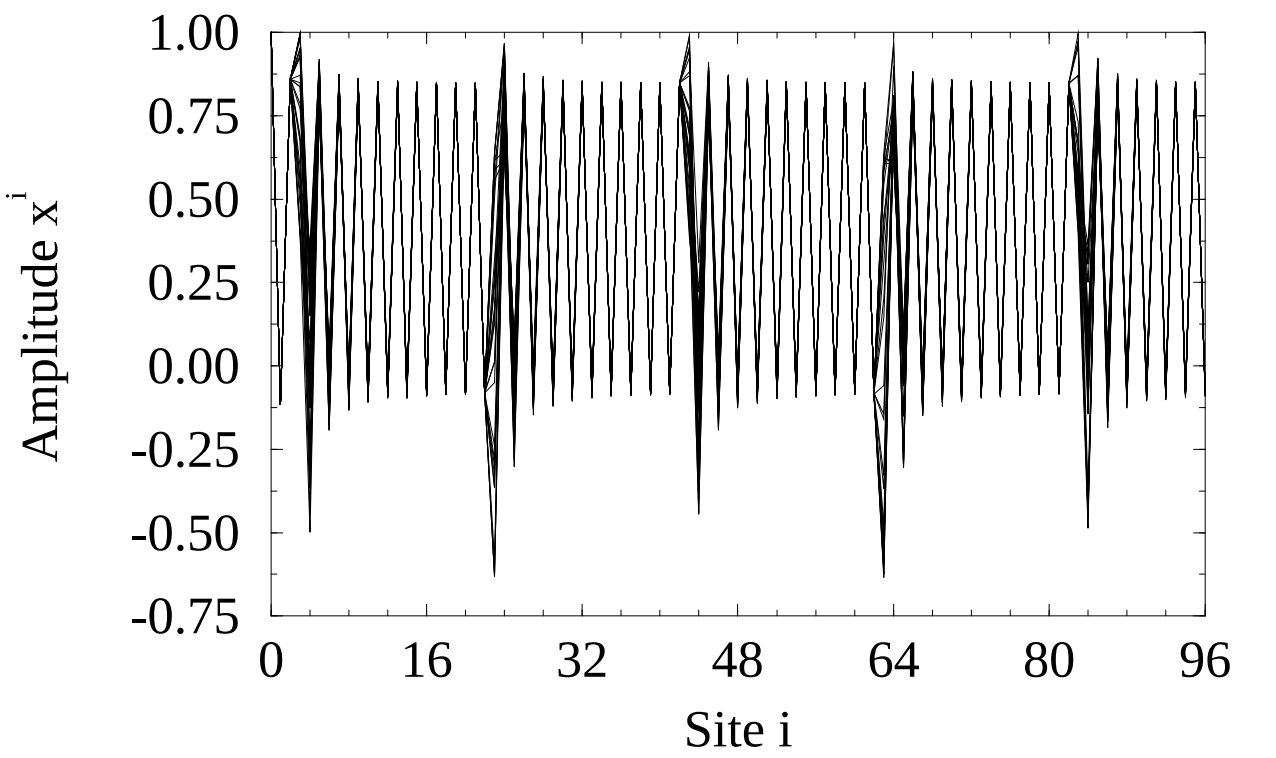

FIG. 2. Self-organized defect lattice in which the distance between defects is 20 lattice sites. The nonlinearity is $\alpha = 1.7$ and the coupling strength is $\epsilon = 0.11595$.

We would like to stress here that this periodicity is in the spatial direction, and that the periodicity can be very long. As is shown in Fig. 3, we find that the distance between defects diverges logarithmically when approaching a critical value $\epsilon_c \approx 0.11525$ beyond which the system is attracted to the zigzag solution for all initial conditions. The distance between defects throughout a lattice is nearly constant and increases in basic steps of one when approaching $\epsilon_c$ (from a minimum of 3 sites). Since there are two phases, this yields lattices in which either most of the successive defects are in the same phase, or in which the phases alternate.

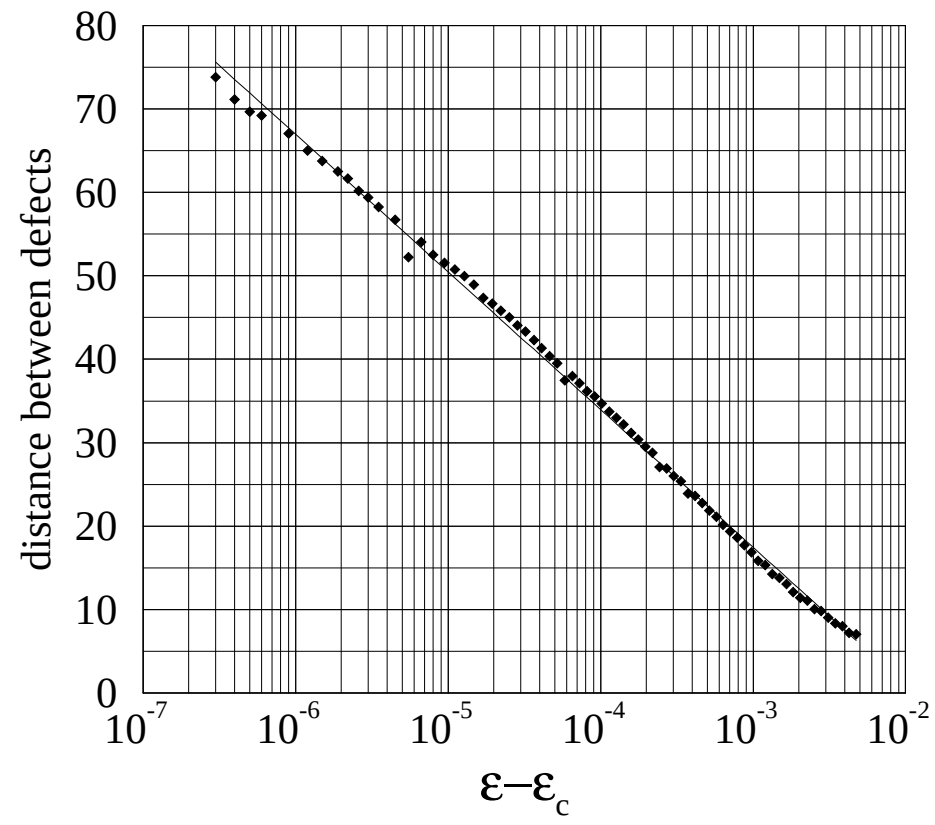

FIG. 3. Distance between successive defects versus $\epsilon - \epsilon_c$ with $\epsilon_c = 0.11525$. The nonlinearity is $\alpha = 1.7$, and 2.5e6 time steps were discarded for each point calculated.

In order to obtain an impression of the extent to which the chaotic motion of the defects is localized, the stationary Lyapunov exponents corresponding to Fig. 2, which can immediately be found as the eigenvalues of the product of Jacobi matrices [4]

$$\lambda^i = \log(1-\epsilon) + \frac{1}{T} \sum_{n=1}^{n=T} \log f'(x_n^i), \tag{3}$$

are depicted in Fig. 4. Due to the upper triangle of the Jacobi matrix being zero for Eq. 1, there is no mixing and the $i$-th exponent represents the local chaoticity of the $i$-th site.

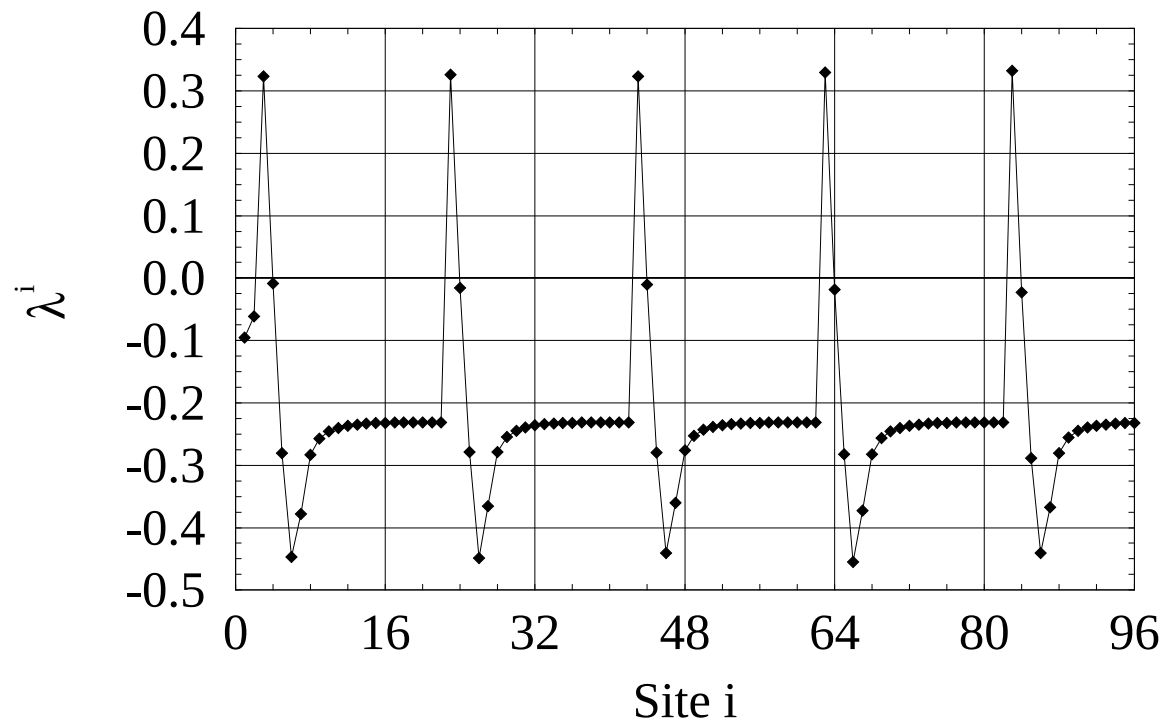

FIG. 4. Stationary Lyapunov exponents corresponding to Fig. 2. The product of 10000 Jacobi matrices was taken. For each defect, there is only one lattice site with a positive Lyapunov exponent. The nonlinearity is $\alpha = 1.7$ and the coupling strength is $\epsilon = 0.11595$.

As can readily be seen from Figs. 2 and 4, the chaotic motion is generally localized at one lattice site, while it is damped in the down-flow direction. The surprising aspect here is that the last site before a new defect shows neither any significant remnant motion, nor a substantial deviation from the zigzag solution which is known to be absolutely stable. That is to say, it is not only stable in the stationary frame, but also in any co-moving frame [6,11].

This is also clearly visible in the spatial return map [2,3] given in Fig. 5, where the straight lines indicate that there is only a very weak correlation between the value of a site prior to the defect and the defective site itself.

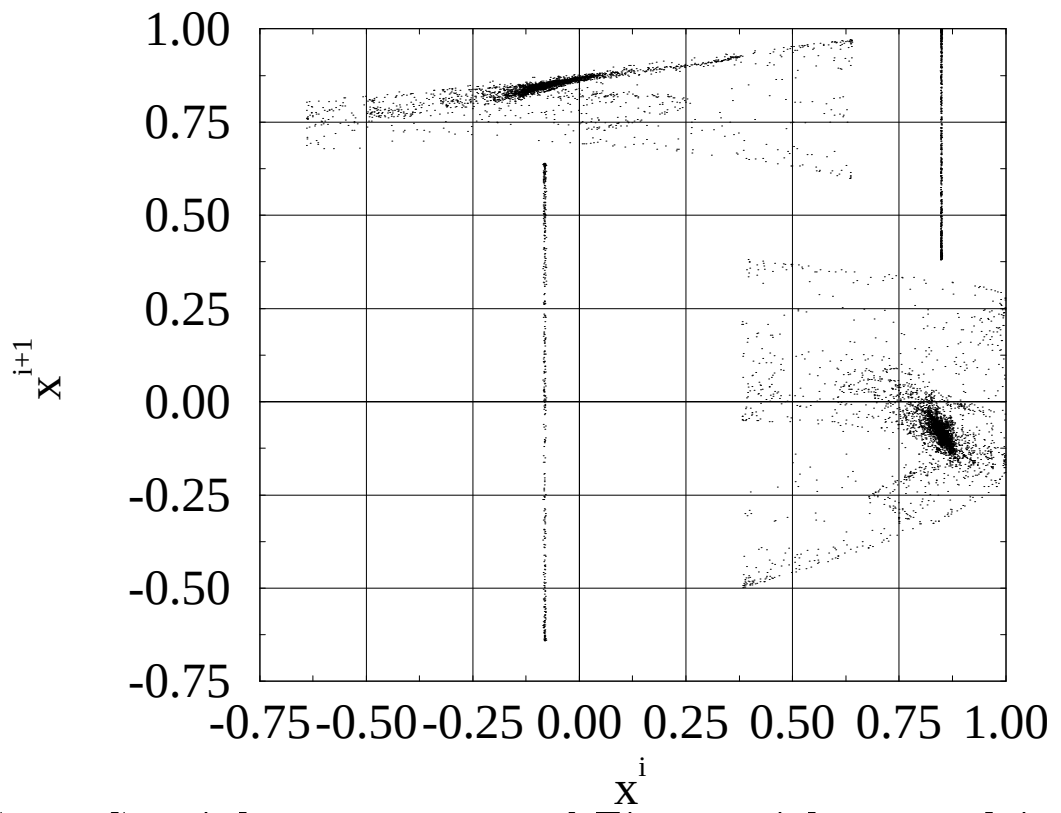

FIG. 5. Spatial return map of Fig. 2 with 50 2nd iterates overlaid. The straight lines clearly indicate that the lattice sites preceding chaotic ones all have (nearly) the same constant amplitudes. The nonlinearity is $\alpha = 1.7$ and the coupling strength is $\epsilon = 0.11595$.

In other words, the zigzag solution is stable to at least several percent of global (and thus also local) noise, but in the defect lattice we first see damping to far within the basin of the stable zigzag pattern and then suddenly another defect at a certain (minimum) distance.

In order to obtain a quantitative estimate on how close to the zigzag pattern the lattice is, we plotted $x^{2i} - x^*$ versus space in Fig. 6 with $i$ shifted such that the defect is at site $i = 1$.

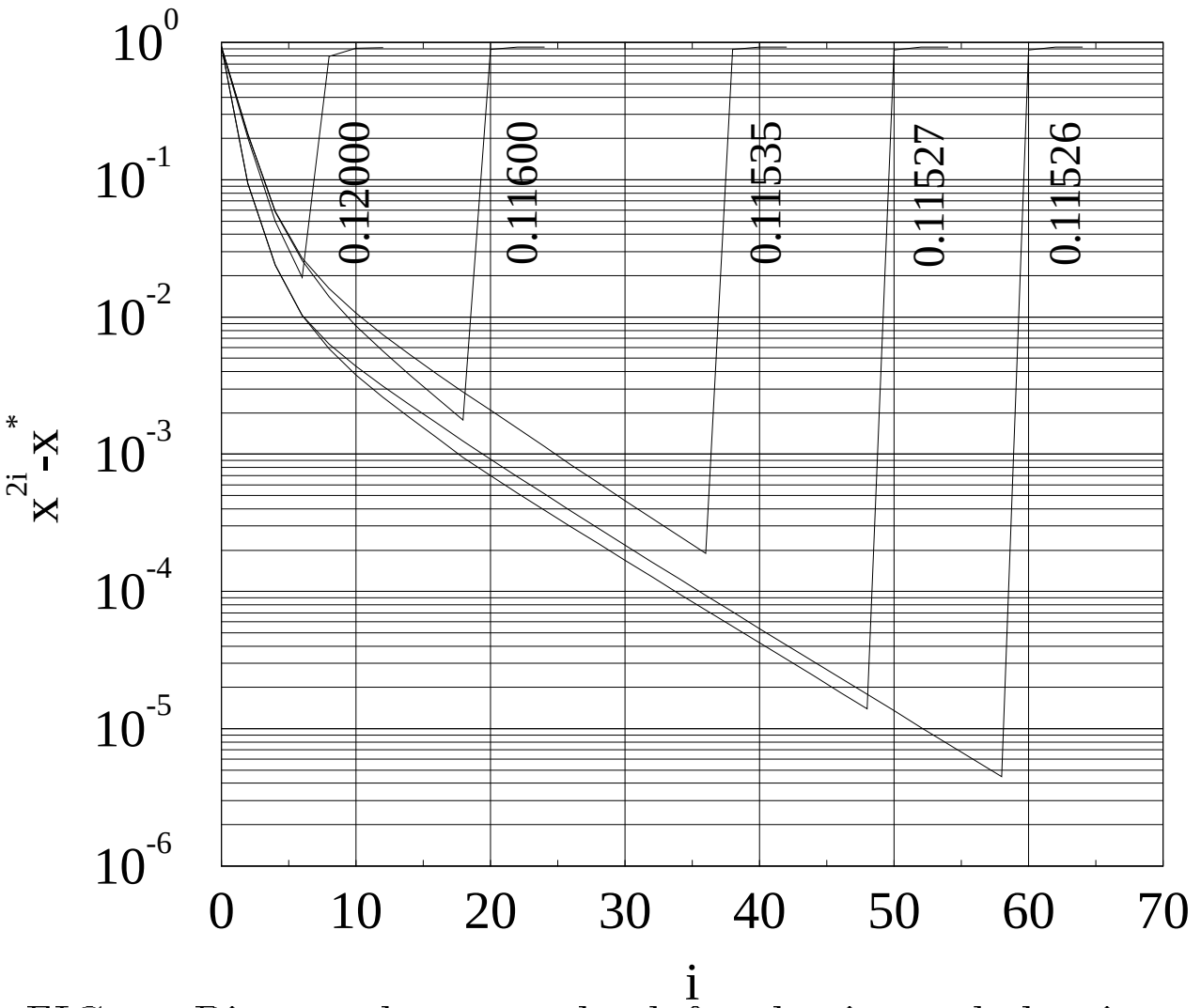

FIG. 6. Distance between the defect lattice and the zigzag pattern. The numbers indicate the values of the coupling constant $\epsilon$, and the nonlinearity is $\alpha = 1.7$.

The average of 1000 samples per every 2nd time step was taken of the sites following the first defect after a spatial transient of 100 sites. What is notable is that when approaching the critical value $\epsilon_c$, we do not see a decay in the damping rate like is case for usual critical phenomena. Instead we see that a perturbation remains decaying exponentially to $x^*$ regardless of the proximity to $\epsilon_c$. Consequently, there must be some threshold related to the phenomena itself which is lowered.

We also found that the distance between defects can be reduced by inserting a small amount of global noise and that for a sufficiently large amount of global noise, the entire lattice falls on the zigzag attractor. Hence the lattice evolves towards a state very near the breakdown of stability for a given distance between defects, and accordingly, we clearly have some kind of self-organization. It should be stressed that Fig. 2 is not a 'lucky' special case, but the general final state for the given parameters when starting from random initial conditions. After the transients have died out, chaos is localized and seems to defy our intuition of chaotic motion countering regularity.

In order to unravel the apparent contradictions (i.e. how is it possible that chaotic defects self-organize into a periodic lattice; how can sites well in the basin of the zigzag attractor be followed by a defect?), we analyze the phenomena described above by employing the fact that the sites just before a defect virtually assume the values of the zigzag pattern. The second iterate of Eq. 1 can then be approximated as

$$x^i_{n+2} = (1-\epsilon) f\left((1-\epsilon) f(x^i_n) + \epsilon f(x^*_2)\right) + f(x^*_1), \quad (4)$$

where $x^*_1$ and $x^*_2$ are the stable zigzag solutions.

This enables us to plot $x^i_{n+2}$ as a function of $x^i_n$ like in Figs. 7 and 8 which depict the situations just before and just after the critical point $\epsilon_c$.

From these figures we can infer that for $\epsilon > \epsilon_c$, there are two distinct basins: one for a chaotic attractor and one for the fixed point of the zigzag pattern, while for $\epsilon < \epsilon_c$ there is only one basin. At $\epsilon_c$ we therefore have a boundary crisis [12] at which the boundary of the chaotic attractor (not the boundary of the basin of the chaotic attractor), in this case given by the second iterate of the origin, touches an unstable fixed point.

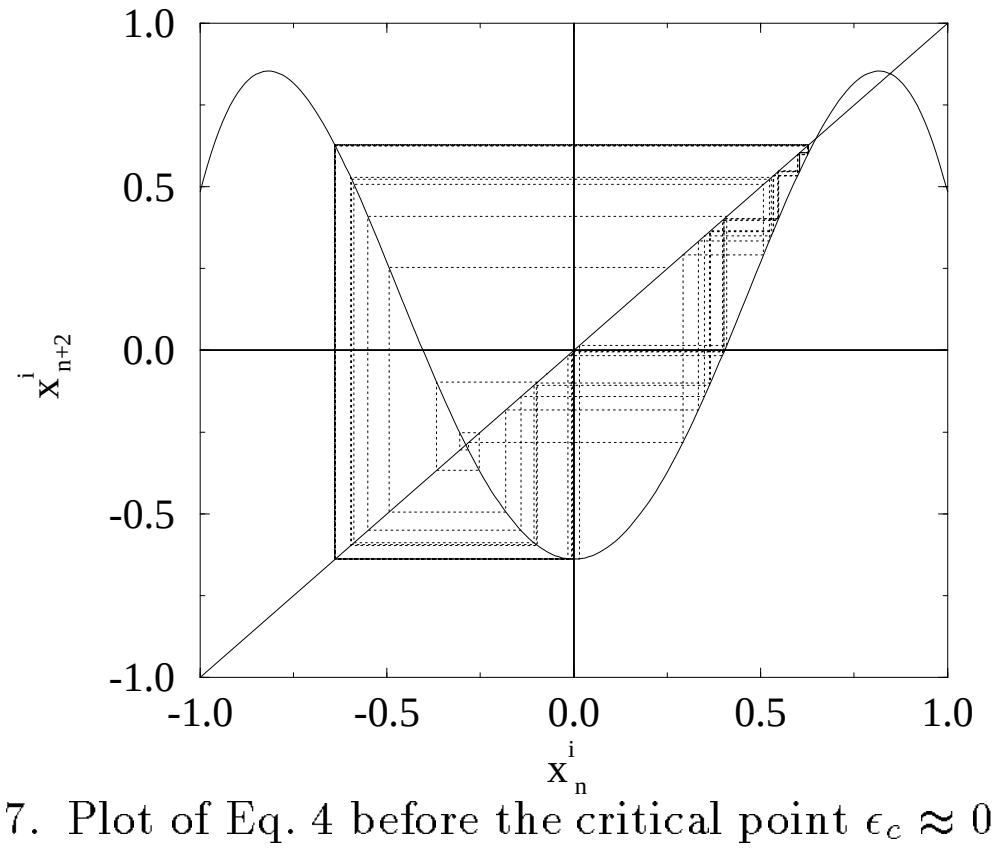

FIG. 7. Plot of Eq. 4 before the critical point $\epsilon_c \approx 0.11525$. The nonlinearity is $\alpha = 1.7$ and the coupling constant is $\epsilon = 0.12$. The dotted line indicates the first 50 iterates of 0.0.

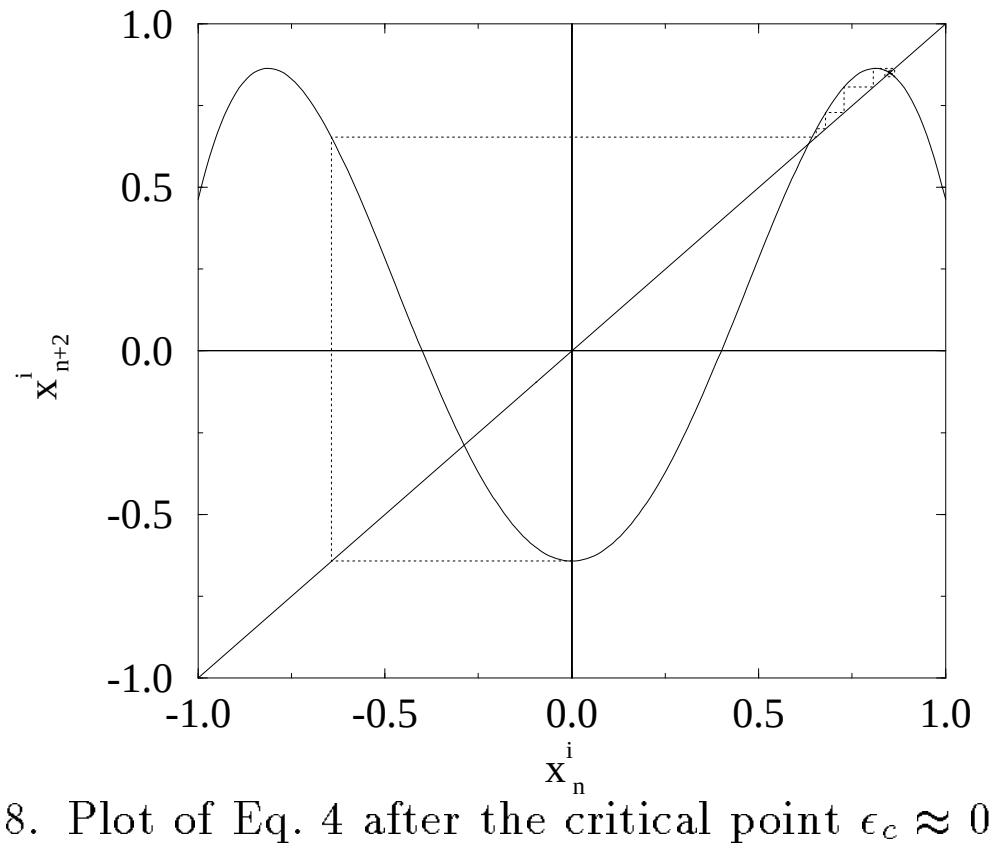

FIG. 8. Plot of Eq. 4 after the critical point $\epsilon_c \approx 0.11525$. The nonlinearity is $\alpha = 1.7$ and the coupling constant is $\epsilon = 0.11$. The dotted line indicates the first 25 iterates of 0.0.

We find that the self-organization of the lattice can be associated with the following scenario:

When starting from random initial conditions, some lattice sites will be attracted to the zigzag attractor. Let us assume $x^{i-1}$ is one such site. The question arising now is under what circumstances will $x^i$ be a defect. If $\epsilon < \epsilon_c$ we have only one basin of attraction, and (eventually) all defects will disappear. If $\epsilon > \epsilon_c$, however, we find two distinct basins, implying that a site in the chaotic basin is (at least initially) a defect.

Since the gap between the chaotic attractor and the basin boundary of the zigzag attractor is quite narrow for $\epsilon > \epsilon_c$, a small modulation of $x^{i-1}$ will create leaks to the zigzag basin. If such a leak exists, $x^i$ cannot be a defect and it will eventually be attracted to the zigzag solution.

The zigzag pattern is a periodic attractor with a negative Lyapunov exponent however. Accordingly, modulation will be damped in the down-flow direction until at some site the modulation is too small to create leaks to the basin of the periodic attractor. If at this stage we have a site in the chaotic basin it will form a defect. When starting from random initial conditions, this will almost always be the case, and we obtain a periodic lattice.

A consequence of the presented scenario is that only the minimum distance between defects is determined. This can easily be verified by chosing a zigzag pattern as initial condition with defects inserted at selected sites. If the distance between two successive defects is larger than the minimum distance, nothing will happen, while if the distance between two successive defects is smaller than the minimally allowed distance, the down-flow defect will be pushed away. It should be mentioned here that it is not possible for a defect to be annihilated when located between (nearly) zigzag sites. This is due to the fact that if the site of a defect leaks to the basin of the periodic attractor, automatically the next down-flow site ends up in the basin of the chaotic attractor.

Since the gap between the edge of the chaotic attractor and the unstable fixed point is approximately proportional to $\epsilon - \epsilon_c$, we have that the threshold $t$ below which the chaotic motion must be damped is also proportional to $\epsilon - \epsilon_c$ as can also immediately be inferred from Fig. 6. The Lyapunov exponent of the zigzag pattern is more or less constant close enough to $\epsilon_c$, implying that $\delta^k \propto exp(-k|\lambda|)$ with $\delta$ the size of a perturbation, and $k$ the distance from the perturbation. Hence the minimum distance $k$ from a perturbation (and thus the minimum distance in the defect lattice) for which we have that $\delta < t$ is given by $k \propto \log(\epsilon - \epsilon_c)$, yielding indeed the observed logarithmic divergence of the distances between defects.

In the present letter we have only studied the fully up-flow coupled case. We have confirmed however that the inclusion of a small down-flow coupling term (i.e. coupling to site $x^{i+1}$) does not affect our results qualitatively. We believe that this is a further indication that our new mechanism of self-organization may be universal and can be encountered elsewhere.

In the diffusively coupled logistic lattice, Brownian motion of defects has been reported [2] in a region of parameter space slightly above the regular zigzag regime. Although we could not observe any defect lattices, we nevertheless believe that the Brownian motion too is the result of a boundary crisis, the only difference with the one-way coupled case being that leaks always exist to some extent allowing the defect to move to the right or left.

It may be interesting to stray somewhat from the main topic and compare our present findings with a standard approach in physics. Let us assume for a moment that Fig. 2 is presented to us as the result of a well executed real-world experiment. It would then be quite natural to attempt an analysis on the basis of attractive and repulsive (quasi-) forces associated with the defects. From our analysis it is clear however that a forces picture would be rather artificial. The point we would like to make here is of course not an attempt to discredit an often extremely successful technique, but that there may very well be cases of actual (likely complex) physical systems for which a dynamical systems approach of analysis is not only more effective but also closer to the underlying nature of the phenomenon, an example of which could be our defect lattice.

In conclusion, we found a new type of self-organized lattice in which chaotic defects occur periodically when starting from random initial conditions. The encountered phenomena constitute a novel mechanism of self-organization which was successfully analyzed and associated with a boundary crises.

**Acknowledgements**

We would like to thank K.Y.M. Wong, T. Yamamoto and N.B. Ouchi for fruitful discussions. This work was supported by a Grant-in-Aid for Scientific Research from the Ministry of Educations, Science and Culture, Japan under grant no. 05836006 and 93043.

---

* e-mail: frederik@complex.c.u-tokyo.ac.jp

† e-mail: kaneko@cyber.c.u-tokyo.ac.jp